\documentclass[showpacs,amsmath,amssymb,aps,10pt,reprint,superscriptaddress,prb]{revtex4-1}
\usepackage{graphicx}
\usepackage{bm}
\usepackage[breaklinks=true,colorlinks=true,linkcolor=blue,urlcolor=blue,citecolor=blue]{hyperref}
\usepackage{graphics}
\usepackage{dcolumn}
\usepackage{amsmath,amssymb}
\usepackage{mathdots}
\usepackage{natbib}
\usepackage{soul,color}
\usepackage{epstopdf}
\usepackage[note-name]{notes2bib}

\begin{document}

\title{Tuning electric charge scattering in YBCO single crystals via irradiation with MeV electrons}

\author{R. V. Vovk}
\author{G. Ya. Khadzhai}
\author{O.~V.~Dobrovolskiy}
    \affiliation{Physikalisches Institut, Goethe University, 60438 Frankfurt am Main, Germany}
    \affiliation{Physics Department, V. Karazin Kharkiv National University, 61077 Kharkiv, Ukraine}

\begin{abstract}
Irradiation with electrons is an efficient approach to inducing a large number of defects with a minimal impact on the material itself. Analysis of the energy transfer from an accelerated particle smashing into the crystal lattice shows that only electrons with MeV energies produce point defects in the form of interstitial ions and vacancies that form perfect scattering centers. Here, we investigate the changes in the resistive characteristics of YBCO single crystals from the 1-2-3 system after several steps of low-temperature irradiation with $0.5-2.5$\,MeV electrons and irradiation doses of up to $8.8\times10^{18}$\,cm$^{-2}$. The penetration depth of such electrons is much larger than the crystal thickness. We reveal that defects appearing in consequence of such electron irradiation not only increase the residual resistance, but they affect the phonon spectrum of the system and lower the superconducting transition temperature linearly with increase of the irradiation dose. Furthermore, the irradiation-induced defects are distributed non-uniformly, that manifests itself via a broadening of the superconducting transition. Interestingly, the excess conductivity remains almost unaffected after such electron irradiation.
\end{abstract}

\maketitle

\section{Introduction}
As is known, electron irradiation along with thermal processing \cite{Jor90pcs,Kha18pcs}, introduction of artificial defects \cite{Lan10inb,Vov15jms} as well as application of a high pressure \cite{Chu87prl,Vov14cap,Fer04prb,Sol16cap,Fan15prb,Sol16phb} and strong magnetic \cite{Bon01ltp,Vov18ltp} or high-power-level electromagnetic \cite{Lar15nsr,Dob18rrl} fields belong to efficient approaches to modify the electro-physical and structural properties of superconducting compounds. In particular, the major experimental problem in studies of the effect of disorder in high-$T_c$ superconductors is the introduction of point defects with a minimal impact on the material itself. In the case of chemical substitutions \cite{Gin89boo,Yan90lcm,Akh02phb,Vov15pcs}, there is always a question of whether the foreign ions change not only the scattering but other parameters, such as the chemical potential and the band structure. Irradiation with energetic particles is an alternative way to introduce defects. However, the nature of the defects produced depends on the irradiation type \cite{Dam63boo}. Protons, $\alpha$ particles, and neutrons most likely produce cascades of clusters of defects, while heavy ions produce columnar tracks or linear defects \cite{Lan10inb,Ghi17nsr}, which are difficult to analyze within simplified pointlike potential scattering models. In addition to area-selective removal of the superconducting material, allowing one to efficiently manipulate magnetic flux lines (Abrikosov vortices) \cite{Bra95rpp,Bae95prl,Har96sci,Cas97apl,Cri05prb,Dob11snm,Zec17pra,Dob17pcs}, irradiation with ions at energies ranging from a few to a few tens of keV leads to surface corrugation, material amorphization, vacancy generation and ion implantation in the processed region \cite{Pau04prb,Kom14apl,Lar14apl,Dob15bjn,Dob15met,Dob17nsr}. In this respect, irradiation with electrons is an efficient approach to inducing a large number of defects \emph{without modification of the composition} of the irradiated sample \cite{Rul03prl,Vov18ssc,Miz14nac,Aza17jms}. Indeed, these defects are charged, but the overall charge change is compensated, so that there is a negligible shift of the chemical potential due to irradiation \cite{Gia92prb}. Analysis of the energy transfer from an accelerated particle smashing into the crystal lattice shows that only electrons at energies of $1-10$\,MeV produce point defects in the form of interstitial ions and vacancies (Frenkel pairs) that presumably form perfect scattering centers \cite{Gia92prb}. According to estimates for Y-Ba-Cu-O, electrons at an energy of $1$\,MeV cause shifts of any of the constituent atoms, while their penetration depth exceeds the crystal thickness which is between $10\,\mu$m and $100\,\mu$m, typically \cite{Gia92prb}. For such crystals, the estimated energy losses for $1$\,MeV electrons amount to $3$\% to $8$\%. As revealed by electron microscopy, electron irradiation leads to the formation of point disorder and/or small clusters which act as pinning sites \cite{Ran99sst}.

While there have been extensive investigations of different aspects of the conducting and superconducting properties of the 1-2-3 YBCO system \cite{Bla94rmp,Vov18ltp}, much less work was concerned with electron-irradiation induced changes in the electric charge scattering therein. In particular, there has been a series of studies of the influence of electron irradiation with energies up to $3$\,MeV and fluences up to $7\times10^{19}$\,cm$^{-2}$ on the superconducting and normal-state transport properties of Y-Ba-Cu-O single crystals, such as the superconducting transition temperature $T_c$, the critical current density $J_c$, the pinning strength for magnetic flux quanta, as well as the low-temperature resistivity \cite{Gia92prb,Dwo94pcs,Rul00epl,Bon01prb,Rul08epl}. At the applications-related facet, an enhancement of the critical current density after electron irradiation is of major importance. This enhancement is caused by irradiation-shifted Cu atoms in the CuO$_2$ planes, which act as strong pinning sites \cite{Gia92prb}. Specifically, irradiation of YBCO single crystals with $3$\,MeV electrons below 10\,K and a fluence of $\simeq5\times10^{19}$\,cm$^{-2}$ results in the appearance of a quasi-two-dimensional system with a characteristic $T_c$ for each of the ``phases'' \cite{Dwo94pcs}. This is accompanied by a decrease of $T_c$ in conjunction with a broadening of the superconducting transition and an increase of the normal-state resistivity \cite{Dwo94pcs}. The irradiation-induced phase segregation appears in consequence of the anisotropy of the damaged areas in the sample produced by irradiation. Namely, oxygen defects primarily appear in the basal Cu(1)-O(4) plane so that the irradiation-induced decrease of $T_c$ in YBCO is associated with the displacement of oxygen and copper in the CuO$_2$ planes as well as with the irradiation-induced point disorder \cite{Rul00epl}.

The strong influence of point defects on the superconducting characteristics of YBCO single crystals is caused by the small coherence length \cite{Wel89prl,Fri89prb,Sol16cap}. These defects  behave as strong pinning sites for magnetic flux quanta \cite{Bla94rmp}. At the same time, point defects noticeably affect the normal-state resistivity of systems with metallic conductivity. This manifests itself via an enhancement of the residual resistivity along with a change in the system's phonon spectrum \cite{Kag66etp}. Accordingly, investigations of the influence of electron irradiation on the electrical resistance of YBCO single crystals are expected to yield important information on the interaction of the charge carriers with the phonon and defect subsystems. Here, we show that defects appearing in consequence of electron irradiation not only increase the residual resistance, but they also affect the phonon spectrum of the system and lower the superconducting transition temperature. Furthermore, the irradiation-induced defects are distributed non-uniformly, that manifests itself via a broadening of the superconducting transition. Interestingly, the excess conductivity remains almost unaffected after the used electron irradiation.

\section{Experiment}
The samples are YBCO single crystals grown in a gold crucible by the solution-melt technique \cite{Obo06ltp}. After the growth, the crystals were saturated in an oxygen atmosphere at $430^\circ$ for four days. All investigated samples were twinned, while the twin planes had a block structure. The electrical resistance was measured in the standard four-probe geometry. The typical dimensions of the crystals were $(1.5-2)\times(0.2-0.3)\times(0.01-0.02)$\,mm$^3$, where the smallest size corresponds to the $c$-axis. The transport current was applied along the largest side of the sample. The distance between the voltage contacts was $1$\,mm. Electron irradiation was done with electrons at energies in the range $0.5-2.5$\,MeV in a cryostat at $T\simeq10$\,K. The cumulative dose $10^{18}$\,cm$^{-2}$ at an electron energy of $2.5$\,MeV produces a concentration of $10^{-4}$ displacements per atom, averaged over all sublattices \cite{Gia92prb}. The helium cryostat allowed for measurements in the temperature range $10$\,K $< T < 300$\,K directly after consequent irradiation steps. While measurements were done on five YBCO single crystals, in what follows we discuss the data acquired on one exemplary sample, since the scattering of the data and deduced parameters is less than 5\% for different samples.

\section{Results and discussion}

\begin{figure}[t!]
    \centering
    \includegraphics[width=0.85\linewidth]{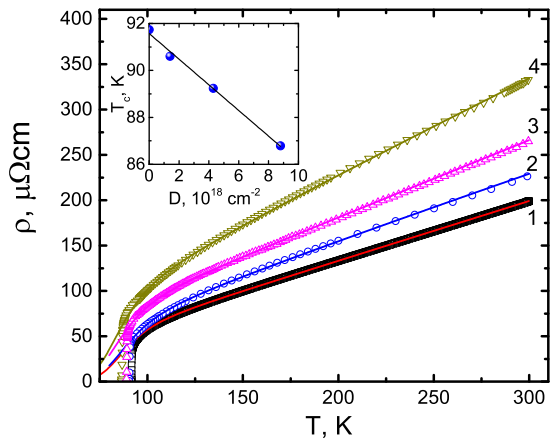}
    \caption{Temperature dependences of the normal-state resistivity for a series of irradiation doses. Symbols: experiment. Lines: fits to Eqs. (\ref{e1}) and (\ref{e2}). Curve numbering: Before irradiation (1), $D =1.4\times10^{18}$\,cm$^2$ (2), $D =4.3\times10^{18}$\,cm$^2$ (3), and $D =8.8\times10^{18}$\,cm$^2$ (4).  Inset: Dependence of $T_c$ on the irradiation dose $D$.}
    \label{f1}
\end{figure}

\begin{figure}[t!]
    \centering
    \includegraphics[width=0.72\linewidth]{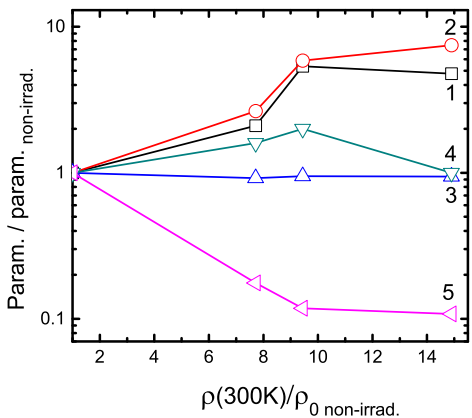}
    \caption{Dependences of the relative changes of the fitting parameters to Eqs. (\ref{e1}) and (\ref{e2}), and $RRR = \rho(\mathrm{300\,K})/\rho_0$ on the  normalized residual resistivity in the non-irradiated state. 1 --- $\theta/\theta_\mathrm{non-irrad.}$, 2 --- $C_3/C_{3\,\mathrm{non-irrad.}}$, 3 --- $ T_1/T_{1\,_\mathrm{non-irrad.}}$, 4 --- $b/b_\mathrm{non-irrad.}$, and 5 --- $RRR$.}
    \label{f2}
\end{figure}

The temperature dependences of the normal-state resistivity $\rho_{nab}(T)$ are presented in Fig. \ref{f1} for a series of irradiation doses. The $\rho_{nab}(T)$ curves fit to
\begin{equation}
    \label{e1}
    \rho_{nab}(T) = \frac{1}{\frac{1}{\rho_0 + \rho_{ph}} + b(e^{T_1/T} - 1)},
\end{equation}
where $\rho_0$ is the residual resistivity due to charge carriers scattering on defects and $\rho_{ph}$ is due to scattering on phonons, described by the Bloch-Gr\"uneisen formula \cite{Col65jap}
\begin{equation}
    \label{e2}
    \rho_{ph} = C_3 \left(\frac{T}{\Theta}\right)^3 \int_0^{\theta/T}\frac{e^x x^3 dx}{(e^x - 1)^2}.
\end{equation}
In Eq. (\ref{e1}), the term $b(e^{T_1/T} -1)$ describes some excess conductivity. The fits $\rho_{nab}(T)$ to Eqs. (\ref{e1}) and (\ref{e2}) are shown in Fig. \ref{f1} by solid lines. In the initial, non-irradiated state the fitting parameters are the residual resistivity $\rho_0 =1.95\,\mu\Omega$cm, the Debye temperature $\theta = 41.5$\,K, the phonon scattering coefficient $C_3 = 54.62\,\mu\Omega$cm, $T_1 = 1132$\,K, and $b = 3.2\times10^{-8}$\,($\mu\Omega$cm)$^{-1}$. The sample's resistivity at $300$\,K is $\rho(300\,\mathrm{K}) =199\,\mu\Omega$cm and it is characterized by $RRR = 102$, as defined in the next paragraph. The fitting error does not exceed 1\%. The evolution of the fitting parameters after consequential irradiation steps is presented in Fig. \ref{f2}.

Proceeding to an analysis of the effect of the low-temperature irradiation at $0.5-2.5$\,MeV energies on the resistive characteristics of the crystals, we begin with a general remark that the accumulation of defects in the sample due to the consequential electron irradiation is characterized by the residual resistivity $\rho_0$. The evolution of $\rho_0$ with increase of the irradiation dose is non-monotonic and it has a tendency to increase. This implies both, an accumulation of defects in consequence of the irradiation and annealing of the sample during the temperature sweep to 300\,K. We note that the residual resistivity ratio, $RRR = \rho_{300\,\mathrm{K}}/\rho_0\approx (\rho_{ph}+\rho_0)/\rho_0$, which is a measure of the disorder degree in the sample, decreases from $RRR\sim100$ ($\rho_{ph} \gg\rho_0$) to $RRR\sim10$ ($\rho_{ph} \geq\rho_0$) already after the first irradiation dose. By contrast, after further irradiation steps $RRR$ increases weakly.

The fitting parameters to Eqs. (\ref{e1}) and (\ref{e2}) allow us to analyze their evolution as a function of $\rho_0$ in Fig. \ref{f2}. Specifically, the increasing disorder degree, as reflected in the increase of $\rho_0$, causes a five-fold increase of the Debye temperature $\theta$. We attribute the small value of $\theta$ to the anisotropy of our samples as the interaction between the layers is much weaker than that in the layer planes. Thus, $\theta$  associated with the transverse oscillations (along the c-axis) is much smaller than $\theta$ related to the transverse oscillations in the layer planes \cite{Ans88etp}. Since $\Delta\theta/\theta \approx -\alpha\Delta V/V + \beta\Delta f/ f$, the larger $\Delta\theta/\theta$ ratio is stipulated by a larger $\Delta f/f$ ratio. Here, $\Delta V$ is the change of the unit cell volume while $\Delta f$ is the force constant change. In this way, irradiation-induced defects effectively increase the interaction between the layers. This is reflected in the isotropization of the phonon spectrum and results in the enhancement of the Debye temperature. Given the stoichiometry of the sample, we come with $\langle\theta\rangle\approx345$\,K for the Debye temperature averaged over all elements.

The parameter $C_3$ characterizing the phonon scattering intensity of the charge carriers increases with increase of the irradiation dose. This agrees with the data of Ref. \cite{Kho83fnt} for transient metals and is likely associated with the phonon spectrum deformation \cite{Kag66etp}. The parameters describing the excess conductivity, $T_1$ and $b$, are almost constant, namely $T_1 \approx(1080\pm30)$\,K and $b\approx (4\pm1)\times10^{-8}$\, $\mu\Omega$m$^{-1}$cm$^{-1}$. We note that $T_1$ is close to the pseudogap value deduced in Ref. \cite{Aza17jms} before the irradiation. Here, specific mechanisms of quasiparticle scattering \cite{Apa02prb65,Vov03prb,Ada94ltp,Vov03prl,Cur11prb} may play a role.

We turn to the superconducting characteristics of the samples with $T_c$ being determined at the low-temperature maximum in the derivative $d\rho/dT$. The dependence of $T_c$ on the irradiation dose $D$ qualitatively agrees with Ref. \cite{Bon01prb} and it fits to the law $T_c(D) = (91.6\pm0.1\,\mathrm{K}) - (0.55\pm0.03\,\mathrm{K})\times D$, as shown in the inset of Fig. \ref{f1}. The linear law  means that the defects responsible for the decrease of $T_c$ are not annealed at $T\leq 300$\,K. These defects are non-magnetic interstitial atoms shifted from their regular positions by incident electrons \cite{Gia92prb,Bon01prb,Abr60etp}.
\begin{figure}[t!]
    \centering
    \includegraphics[width=0.8\linewidth]{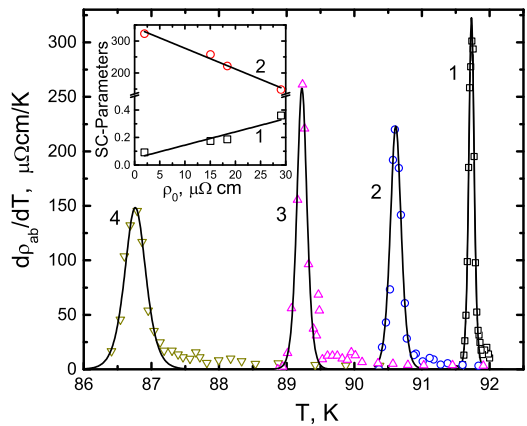}
    \caption{Derivatives $d\rho/dT$ near $T_c$. Inset: widths (1) and heights (2) of the $d\rho/dT$ maxima in dependence on $\rho_0$.}
    \label{f3}
\end{figure}

Figure \ref{f3} displays the derivatives $d\rho/dT$ near $T_c$. The curves fit to \cite{Rol83boo}
\begin{equation}
    \label{e3}
    \frac{d\rho(T)}{dT} = \frac{\rho_1 e^{-Z}}{w(1 + e^{-Z})^2},
\end{equation}
where $Z = (T_c- T)/w$ and $w$ characterizes the width of the superconducting transition and $\rho_1 \approx 4 w (d\rho/dT)_{max}$.

In Fig. \ref{f3} one sees that the maxima in $d\rho/dT$ shift towards lower temperatures with increase of the irradiation dose. At the same time, the peaks are broadening, while their symmetry is maintained. The latter means that the distribution of the defects after the irradiation remains homogenous within the layer planes. The inset to Fig. \ref{f3} depicts the dependences of the half-height width $\Delta T_{c0.5} \approx 3.52 w$ of the derivative $d\rho/dT$ and the maximum heights on the residual resistivity $\rho_0$. One sees that these characteristics monotonically change with increase of $\rho_0$, that is with increase of the number of defects.

We believe that the preservation of the symmetry of $d\rho/dT$ with increase of $\rho_0$ points to that these are mesoscopic variations of the sample composition which are responsible for the broadening of its superconducting transition. In particular, the superconducting transition broadening means that different regions exist in the sample, whose $T_c$s are in the range $T_c- \Delta T_{c0.5} \leq T\leq T_c+ \Delta T_{c0.5}$. The presence of these regions is associated with the variation of the defect concentration, primarily, of oxygen vacancies. In particular, in the case of smeared ferroelectric phase transitions such regions are known as K\"anzig regions \cite{Rol83boo}. The size of such regions can be estimated as $\simeq10^{-5}-10^{-6}\,$cm. The vanish of the electrical resistance in the sample is caused by the appearance of a superconducting region spreading over the sample, which shunts other regions with a lower $T_c$ as well as regions remaining in the normal state.

\section{Conclusion}
To sum up, we have investigated the effect of low-temperature irradiation with electrons at energies of $0.5-2.5$\,MeV on the resistive properties of YBCO single crystals. Namely, such irradiation has been reveled to lead to the appearance of a larger number of defects already at doses of $\approx 1.4\times10^{18}$\,cm$^{-2}$ (Fig. \ref{f1}, curve 2). With a further increase of the irradiation dose the disorder degree increases slowly (Fig. \ref{f2}, curve 5). Irradiation-induced defects reduce the anisotropy of the sample. They enhance charge scattering on phonons, reduce the critical temperature and broaden the superconducting transition. The excess conductivity remains almost unaffected after the used electron irradiation.

Research leading to these results received funding from the European Commission in the framework of the program Marie Sklodowska-Curie Actions --- Research and Innovation Staff Exchange (MSCA-RISE) under Grant Agreement No. 644348 (MagIC).


\begin{thebibliography}{10}
\expandafter\ifx\csname url\endcsname\relax
  \def\url#1{\texttt{#1}}\fi
\expandafter\ifx\csname urlprefix\endcsname\relax\def\urlprefix{URL }\fi
\expandafter\ifx\csname href\endcsname\relax
  \def\href#1#2{#2} \def\path#1{#1}\fi

\bibitem{Jor90pcs}
J.~D. Jorgensen, S.~Pei, P.~Lightfoor, H.~Shi, A.~P. Paulikas, B.~W. Veal,
  \href{http://www.sciencedirect.com/science/article/pii/0921453490906766}{Physica
  C} 167 (1990) 571--578.

\bibitem{Kha18pcs}
G.~Y. Khadzhai, R.~V. Vovk, N.~R. Vovk, S.~N. Kamchatnaya, O.~V. Dobrovolskiy,
  \href{http://www.sciencedirect.com/science/article/pii/S0921453417305051}{Physica
  C} 545 (2018) 14--17.

\bibitem{Lan10inb}
W.~Lang, J.~D. Pedaring, Ion Irradiation of High-Temperature Superconductors
  and Its Application for Nanopatterning, Springer Heidelberg, 2010, Ch.~3, pp.
  81--104.

\bibitem{Vov15jms}
R.~Vovk, G.~Khadzhai, O.~Dobrovolskiy, N.~Vovk, Z.~Nazyrov,
  \href{http://dx.doi.org/10.1007/s10854-014-2558-y}{J. Mater. Sci: Mater.
  Electron.} 26 (2015) 1435--1440.

\bibitem{Chu87prl}
C.~W. Chu, P.~H. Hor, R.~L. Meng, L.~Gao, Z.~J. Huang, Y.~Q. Wang,
  \href{http://link.aps.org/doi/10.1103/PhysRevLett.58.405}{Phys. Rev. Lett.}
  58 (1987) 405--407.

\bibitem{Vov14cap}
R.~V. Vovk, N.~R. Vovk, G.~Y. Khadzhai, O.~V. Dobrovolskiy, Z.~F. Nazyrov,
  \href{http://www.sciencedirect.com/science/article/pii/S1567173914003113}{Curr.
  Appl. Phys.} 14 (2014) 1779 -- 1782.

\bibitem{Fer04prb}
L.~M. Ferreira, P.~Pureur, H.~A. Borges, P.~Lejay,
  \href{http://link.aps.org/doi/10.1103/PhysRevB.69.212505}{Phys. Rev. B} 69
  (2004) 212505.

\bibitem{Sol16cap}
A.~L. Solovjov, L.~V. Omelchenko, R.~V. Vovk, O.~V. Dobrovolskiy, S.~N.
  Kamchatnaya, D.~M. Sergeev,
  \href{http://www.sciencedirect.com/science/article/pii/S1567173916301274}{Curr.
  Appl. Phys.} 16 (2016) 931 -- 938.

\bibitem{Fan15prb}
Y.~Fang, D.~Yazici, B.~D. White, M.~B. Maple,
  \href{http://link.aps.org/doi/10.1103/PhysRevB.92.094507}{Phys. Rev. B} 92
  (2015) 094507.

\bibitem{Sol16phb}
A.~L. Solovjov, L.~V. Omelchenko, R.~V. Vovk, O.~V. Dobrovolskiy, Z.~Nazyrov,
  S.~Kamchatnaya, D.~Sergeyev,
  \href{http://www.sciencedirect.com/science/article/pii/S0921452616301351}{Physica
  B} 493 (2016) 58 -- 67.

\bibitem{Bon01ltp}
A.~V. Bondarenko, A.~A. Prodan, M.~A. Obolenskii, R.~V. Vovk, T.~R. Arouri,
  \href{http://scitation.aip.org/content/aip/journal/ltp/27/5/10.1063/1.1374717}{Low
  Temp. Phys.} 27 (2001) 339--344.

\bibitem{Vov18ltp}
R.~V. Vovk, A.~L. Solovjov, \href{https://doi.org/10.1063/1.5020905}{Low Temp.
  Phys.} 44 (2018) 81--113.

\bibitem{Lar15nsr}
A.~Lara, F.~G. Aliev, A.~V. Silhanek, V.~V. Moshchalkov,
  \href{http://dx.doi.org/10.1038/srep09187}{Sci. Rep.} 5 (2015) 9187.

\bibitem{Dob18rrl}
O.~V. Dobrovolskiy, R.~Sachser, V.~M. Bevz, A.~Lara, F.~G. Aliev, V.~A.
  Shklovskij, A.~I. Bezuglyj, R.~V. Vovk, M.~Huth,
  \href{https://onlinelibrary.wiley.com/doi/abs/10.1002/pssr.201800223}{Rapid
  Res. Lett.} 0  1800223.

\bibitem{Gin89boo}
D.~M. Ginsberg (Ed.), Physical properties of high temperature superconductors
  I., Word Scientific, Singapore, 1989.

\bibitem{Yan90lcm}
Y.~Yan, M.~Blanchin, G.~Fuchs,
  \href{http://www.sciencedirect.com/science/article/pii/0022508890902178}{J.
  Less Comm. Met.} 164--165 (1990) 215 -- 222.

\bibitem{Akh02phb}
M.~Akhavan,
  \href{http://www.sciencedirect.com/science/article/pii/S0921452602008608}{Physica
  B} 321 (2002) 265--282.

\bibitem{Vov15pcs}
R.~V. Vovk, G.~Y. Khadzhai, O.~V. Dobrovolskiy, Z.~F. Nazyrov, A.~Chroneos,
  \href{http://www.sciencedirect.com/science/article/pii/S0921453415002142}{Physica
  C} 516 (2015) 58 -- 61.

\bibitem{Dam63boo}
A.~C. Damask, G.~J. Dienes, Point Defects in Metals, Gordon \& Breach Science
  Publishers Ltd, London, 1963.

\bibitem{Ghi17nsr}
G.~Ghigo, G.~A. Ummarino, L.~Gozzelino, R.~Gerbaldo, F.~Laviano, D.~Torsello,
  T.~Tamegai, \href{https://doi.org/10.1038/s41598-017-13303-5}{Sci. Rep.} 7
  (2017) 13029.

\bibitem{Bra95rpp}
E.~H. Brandt, \href{http://stacks.iop.org/0034-4885/58/i=11/a=003}{Rep. Progr.
  Phys.} 58 (1995) 1465--1594.

\bibitem{Bae95prl}
M.~Baert, V.~V. Metlushko, R.~Jonckheere, V.~V. Moshchalkov, Y.~Bruynseraede,
  \href{https://link.aps.org/doi/10.1103/PhysRevLett.74.3269}{Phys. Rev. Lett.}
  74 (1995) 3269--3272.

\bibitem{Har96sci}
K.~Harada, O.~Kamimura, H.~Kasai, T.~Matsuda, A.~Tonomura, V.~V. Moshchalkov,
  \href{http://www.sciencemag.org/content/274/5290/1167.abstract}{Science} 274
  (1996) 1167--1170.

\bibitem{Cas97apl}
A.~Castellanos, R.~W{\"o}rdenweber, G.~Ockenfuss, A.~v.~d. Hart, K.~Keck,
  \href{https://doi.org/10.1063/1.119701}{Appl. Phys. Lett.} 71 (1997)
  962--964.

\bibitem{Cri05prb}
A.~Crisan, A.~Pross, D.~Cole, S.~J. Bending, R.~W\"ordenweber, P.~Lahl, E.~H.
  Brandt, \href{http://link.aps.org/doi/10.1103/PhysRevB.71.144504}{Phys. Rev.
  B} 71 (2005) 144504--1--10.

\bibitem{Dob11snm}
O.~V. Dobrovolskiy, M.~Huth, V.~A. Shklovskij,
  \href{http://dx.doi.org/10.1007/s10948-010-1055-7}{J. Supercond. Nov.
  Magnet.} 24 (2011) 375--380.

\bibitem{Zec17pra}
G.~Zechner, F.~Jausner, L.~T. Haag, W.~Lang, M.~Dosmailov, M.~A. Bodea, J.~D.
  Pedarnig,
  \href{https://link.aps.org/doi/10.1103/PhysRevApplied.8.014021}{Phys. Rev.
  Applied} 8 (2017) 014021.

\bibitem{Dob17pcs}
O.~V. Dobrovolskiy,
  \href{http://www.sciencedirect.com/science/article/pii/S092145341630096X}{Physica
  C} 533 (2017) 80--90.

\bibitem{Pau04prb}
A.~Pautrat, J.~Scola, C.~Goupil, C.~Simon, C.~Villard, B.~Domeng\`es, Y.~Simon,
  C.~Guilpin, L.~M\'echin,
  \href{http://link.aps.org/doi/10.1103/PhysRevB.69.224504}{Phys. Rev. B} 69
  (2004) 224504--1--5.

\bibitem{Kom14apl}
M.~Kompaniiets, O.~V. Dobrovolskiy, C.~Neetzel, F.~Porrati, J.~Br\"otz,
  W.~Ensinger, M.~Huth, \href{http://dx.doi.org/10.1063/1.4863980}{Appl. Phys.
  Lett.} 104 (2014) 052603.

\bibitem{Lar14apl}
A.~Lara, O.~V. Dobrovolskiy, J.~L. Prieto, M.~Huth, F.~G. Aliev,
  \href{http://scitation.aip.org/content/aip/journal/apl/105/18/10.1063/1.4900789}{Appl.
  Phys. Lett.} 105 (2014) 182402.

\bibitem{Dob15bjn}
O.~V. Dobrovolskiy, M.~Kompaniiets, R.~Sachser, F.~Porrati, C.~Gspan, H.~Plank,
  M.~Huth, \href{http://dx.doi.org/10.3762/bjnano.6.109}{Beilstein J.
  Nanotech.} 6 (2015) 1082--1090.

\bibitem{Dob15met}
O.~V. Dobrovolskiy, M.~Huth, V.~A. Shklovskij,
  \href{http://scitation.aip.org/content/aip/journal/apl/107/16/10.1063/1.4934487}{Appl.
  Phys. Lett.} 107 (2015) 162603--1--5.

\bibitem{Dob17nsr}
O.~V. Dobrovolskiy, M.~Huth, V.~Shklovskij, R.~V. Vovk,
  \href{https://doi.org/10.1038/s41598-017-14232-z}{Sci. Rep.} 7 (2017) 13740.

\bibitem{Rul03prl}
F.~Rullier-Albenque, H.~Alloul, R.~Tourbot,
  \href{https://link.aps.org/doi/10.1103/PhysRevLett.91.047001}{Phys. Rev.
  Lett.} 91 (2003) 047001.

\bibitem{Vov18ssc}
R.~V. Vovk, G.~Y. Khadzhai, O.~V. Dobrovolskiy,
  \href{http://www.sciencedirect.com/science/article/pii/S0038109818302345}{Solid
  State Commun.} 282 (2018) 5 -- 8.

\bibitem{Miz14nac}
Y.~Mizukami, M.~Konczykowski, Y.~Kawamoto, S.~Kurata, S.~Kasahara,
  K.~Hashimoto, V.~Mishra, A.~Kreisel, Y.~Wang, P.~J. Hirschfeld, Y.~Matsuda,
  T.~Shibauchi, \href{http://dx.doi.org/10.1038/ncomms6657}{Nat. Commun.} 5
  (2014) 5657.

\bibitem{Aza17jms}
N.~A. Azarenkov, V.~N. Voevodin, R.~V. Vovk, G.~Y. Khadzhai, S.~V. Lebedev,
  V.~V. Sklyar, S.~N. Kamchatnaya, O.~V. Dobrovolskiy,
  \href{https://doi.org/10.1007/s10854-017-7483-4}{J. Mater. Sci.: Mater.
  Electron.} 28 (2017) 15886--15890.

\bibitem{Gia92prb}
J.~Giapintzakis, W.~C. Lee, J.~P. Rice, D.~M. Ginsberg, I.~M. Robertson,
  R.~Wheeler, M.~A. Kirk, M.-O. Ruault,
  \href{https://link.aps.org/doi/10.1103/PhysRevB.45.10677}{Phys. Rev. B} 45
  (1992) 10677--10683.

\bibitem{Ran99sst}
R.~Rangel, D.~Galvan, G.~Hirata, E.~Adem, F.~Morales, M.~Maple,
  \href{http://stacks.iop.org/0953-2048/12/i=5/a=005}{Supercond. Sci. Technol.}
  12 (1999) 264.

\bibitem{Bla94rmp}
G.~Blatter, M.~V. Feigel'man, V.~B. Geshkenbein, A.~I. Larkin, V.~M. Vinokur,
  \href{http://link.aps.org/doi/10.1103/RevModPhys.66.1125}{Rev. Mod. Phys.} 66
  (1994) 1125--1388.

\bibitem{Dwo94pcs}
F.~Dworschak, U.~Dedek, Y.~Petrusenko,
  \href{http://www.sciencedirect.com/science/article/pii/0921453494918961}{Physica
  C} 235-240 (1994) 1343 -- 1344.

\bibitem{Rul00epl}
F.~Rullier-Albenque, P.~A. Vieillefond, H.~Alloul, A.~W. Tyler, P.~Lejay, J.~F.
  Marucco, \href{http://stacks.iop.org/0295-5075/50/i=1/a=081}{Eur. Phys.
  Lett.} 50 (2000) 81.

\bibitem{Bon01prb}
A.~V. Bondarenko, A.~A. Prodan, Y.~T. Petrusenko, V.~N. Borisenko,
  F.~Dworschak, U.~Dedek,
  \href{https://link.aps.org/doi/10.1103/PhysRevB.64.092513}{Phys. Rev. B} 64
  (2001) 092513.

\bibitem{Rul08epl}
F.~Rullier-Albenque, H.~Alloul, F.~Balakirev, C.~Proust,
  \href{http://stacks.iop.org/0295-5075/81/i=3/a=37008}{Eur. Phys. Lett.} 81
  (2008) 37008.

\bibitem{Wel89prl}
U.~Welp, W.~K. Kwok, G.~W. Crabtree, K.~G. Vandervoort, J.~Z. Liu,
  \href{https://link.aps.org/doi/10.1103/PhysRevLett.62.1908}{Phys. Rev. Lett.}
  62 (1989) 1908--1911.

\bibitem{Fri89prb}
T.~A. Friedmann, J.~P. Rice, J.~Giapintzakis, D.~M. Ginsberg,
  \href{http://link.aps.org/doi/10.1103/PhysRevB.39.4258}{Phys. Rev. B} 39
  (1989) 4258--4266.

\bibitem{Kag66etp}
Y.~M. Kagan, M.~P. Gernov, J. Exp. Theor. Phys. 50 (1966) 1107.

\bibitem{Obo06ltp}
M.~A. Obolenskii, R.~V. Vovk, A.~V. Bondarenko, N.~N. Chebotaev,
  \href{http://link.aip.org/link/?LTP/32/571/1}{Low Temp. Phys.} 32 (2006)
  571--575.

\bibitem{Col65jap}
L.~Colquitt,
  \href{http://scitation.aip.org/content/aip/journal/jap/36/8/10.1063/1.1714510}{J.
  Appl. Phys.} 36 (1965) 2454--2458.

\bibitem{Ans88etp}
N.~V. Anshukova, \emph{et al}, JETP Lett. 48 (1988) 152--154.

\bibitem{Kho83fnt}
V.~I. Khotkevich, B.~A. Merisov, M.~A. Ermolaev, A.~V. Krasnokutskiy, Fiz.
  Nizk. Temp. 9 (1983) 1056.

\bibitem{Apa02prb65}
V.~M. Apalkov, M.~E. Portnoi,
  \href{http://link.aps.org/doi/10.1103/PhysRevB.65.125310}{Phys. Rev. B} 65
  (2002) 125310.

\bibitem{Vov03prb}
R.~V. Vovk, C.~D.~H. Williams, A.~F.~G. Wyatt,
  \href{https://link.aps.org/doi/10.1103/PhysRevB.68.134508}{Phys. Rev. B} 68
  (2003) 134508.

\bibitem{Ada94ltp}
I.~N. Adamenko, K.~E. Nemchenko, V.~I. Tsyganok, A.~I. Chervanev,
  \href{http://link.aip.org/link/?LTP/20/498/1}{Low Temp. Phys.} 20 (1994)
  498--504.

\bibitem{Vov03prl}
R.~V. Vovk, C.~D.~H. Williams, A.~F.~G. Wyatt,
  \href{https://link.aps.org/doi/10.1103/PhysRevLett.91.235302}{Phys. Rev.
  Lett.} 91 (2003) 235302.

\bibitem{Cur11prb}
P.~J. Curran, V.~V. Khotkevych, S.~J. Bending, A.~S. Gibbs, S.~L. Lee, A.~P.
  Mackenzie, \href{http://link.aps.org/doi/10.1103/PhysRevB.84.104507}{Phys.
  Rev. B} 84 (2011) 104507.

\bibitem{Abr60etp}
A.~A. Abrikosov, L.~P. Gorkov, J. Exp. Theor. Phys. 39 (1960) 1781.

\bibitem{Rol83boo}
B.~N. Rolov, V.~E. Yurkevich, Physics of smeared phase transitions, RGU,
  Rostov-on-Don, 1983.

\end{thebibliography}

\end{document}